# Intelligent Message Behavioral Identification System

**Yuvaraju Chinnam** [1], **Bosubabu Sambana** [2]

[1] Department of Computer Science and Engineering,
St. Peter's Engineering College (Autonomous), Hyderabad
chinnamyuvraj@gmail.com

[2] Department of Computer Science and Engineering,
Lendi Institute of Engineering and Technology (A),
JNTU GV, Vizianagaram, Andhra Pradesh, India.
bosukalam@gmail.com

**Abstract:** On social media platforms, the act of predicting reposting is seen as a challenging issue related to Short Message Services (SMS). This study examines the issue of predicting picture reposting in SMS and forecasts users' behavior in sharing photographs on Twitter. Several research vary. The paper introduces a network called Image Retweet Modeling (IRM) that models heterogeneous image retransmission. It considers the user's previous reposting of the image tweet, the next contact in the SMS, and the preferences of the reposted person. Three aspects connected to content. A text-guided multimodal neural network is developed to create a novel multi-faceted attention ranking network methodology. This allows for learning the joint image Twitter representation and user preference representation in the prediction job. Multiple experiments conducted on extensive data sets demonstrate that our approach outperforms current methods on Social Network platforms.

## 1. INTRODUCTION

Today, a Weibo service system similar to Twitter has become a powerful Social media platform for consumers or Customers, who obey other users referred to as "followers" of their posts or comments. Those who are followed to be referred to as "followers." platform is not a critical mechanism. It is a user-friendly environment with the secure platform provided by the social media network channels in the SMS forwarding feature, along with the problem is to create a user or authenticated person. "are followers - followers' tweets forwarded this link to share tweets behavioral model, which is in [1-2] worldwide has attracted widespread attention with curiosity approaching to feed who are rally share their views, comments, and posts. Existing retweet prediction methods [ 1, 3 ] can learn user preference models from text tweets that users reposted to make predictions, but they are limited to text tweets. With the popularity of mobile devices or any PDA through users generate image tweets.

The number has increased dramatically, and 17.2% of Twitter content in Twitter today is related to images [ 2 ]. As a result, it is critical to investigate the image forwarding prediction problem in social media sites. A simple image feed forwarding from one source to various destinations through predefined network channels and a predictable source mechanism to different authenticated user locations is shown in Figure 1. But because image twitter has no distinguishing feature representation [ 2 ], and the SMS or Post comments data is sparse, the existing forwarding prediction methods are far away and very useful for less the prediction of image forwarding problems, recognizing identified issues solve to the proposed new mechanism.

At present, the existing forwarding prediction methods [ 1, 3 ] mainly involve the selection and representation of social media information, including Twitter images and titles, images and users' social roles [ 3 ], and their powerful emotions on posted or seen images and test message[ 4 ]. In recent years, pre-training has been used to achieve good results in various visual recognition tasks using decent convolutional neural networks (Convolutional Neural Networks, CNN) characterized by obtaining an image of high-level visual features [ 5 - 6 ] image push message always visual data, so it is natural.

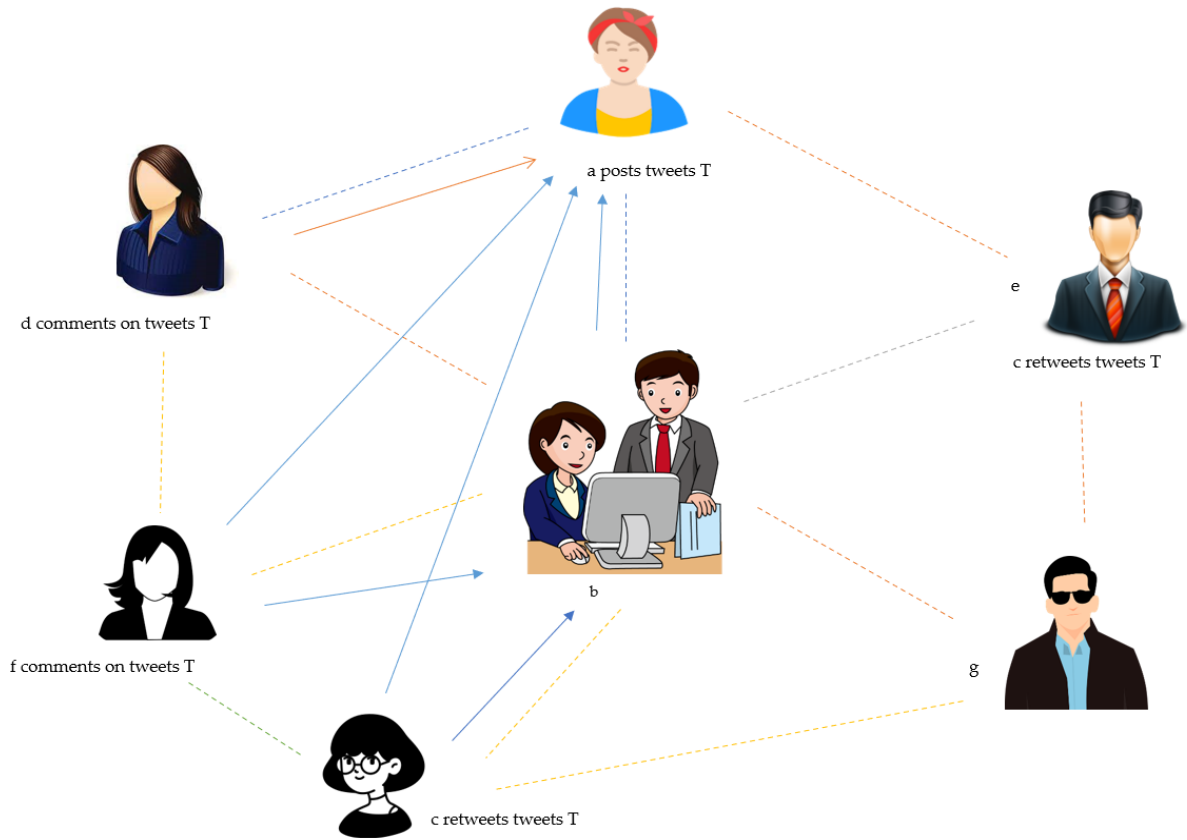

Figure. 1: Intelligent Behavioral Identification and Behavioral Message Communications through allocated channels

Deep convolutional neural networks are used to comprehend the visual representation of picture tweets. [7] Textual, contextual information, such as user comments and headlines, frequently accompanies image tweets. [2]. Background image tweets usually convey Important Information, helpful to understand tweets; we use a Deep Recurrent Neural Network [ 8 ] to learn its semantic representation. Current research uses various models, such as tensor decomposition [9] and Probabilistic Matrix Decomposition [ 10 ]. These models cannot explore the relationship between the image tweet and its title or comment to simulate the user's reposting behavior. We use a Multi-modal Neural Network(MNN) [ 11 ] to learn the Twitter representation of the joint image from the multi-modal content. Thereby, Different forms of complementary information can be provided.

SMS statistics' sparseness is also a complex problem for picture forwarding prediction. In SMS websites, the community between picture tweets and customers is developed via the forwarding relationship of users to photograph tweets. Usually, each user only forwards A small number of image tweets, so the SMS network is sparse. Inspired by the homogeneity hypothesis [ 12 ], we can jointly consider the set information of the user's followers and the user's reposted tweets to solve the sparseness problem of image reposting prediction.

The social impact of forwarding behavior is different between users and other followers. Therefore, we use the attention mechanism [ 13 ] to adaptively integrate user follower preferences to predict target users' image forwarding behavior jointly.

This paper studies the problem of image forwarding prediction from the perspective of multi-modal attention ranking network learning[1]. This paper first proposes a heterogeneous image forwarding model (Image retweet modeling, IRM) network using multi-modal image Twitter. The three aspects of the user's forwarding behavior and tracking relationship predict image forwarding. This paper introduces two sub-network multi-modal neural networks guided by the text. The recurrent neural network learns the semantic representation of the image tweet context information, and the convolutional neural network learns vision Representation.

We use the multi-faceted interest sorting method to build multi-modal neural networks. The metric value of multi-faceted sorting is implicit in the person's desired illustration for image forwarding prediction. The main contributions of this paper are as follows:
   a) Different from previous studies, this paper proposes a heterogeneous IRM network model to simulate the prediction of image forwarding [4].
   b) This model uses the image tweets and related contexts that users forwarded in the past, the user's following relationship in SMS, and user pairs—three aspects of preference information for subsequent content.
   c) A text-guided multi-modal neural network-based attention multi-faceted sorting method is proposed to learn the tracking relationship between user preference representations based on retweets and image tweets prediction.
   d) Use the data set obtained from Twitter to assess the method's performance.
   e) Many experiments show that the method described in this paper outperforms the existing process.

## 2. RELATED WORK

Retweet prediction has been extensively and in-depth researched in recent years. It is a way of information dissemination in today's social media. To accurately model the reposting behavior of users, we divide the current research work into three aspects: User reposting behavior, Feature selection, forwarding model characterization, and user forwarding ranking. We will review related work in these three areas in this section. Feature selection of user reposting behavior: Choosing the relevant factors that affect user reposting behavior. This issue has been deeply studied. Fir daus et al. [4] discussed the influence of user topic-specific emotions on their reposting decision. Experiments show that contextual features affect the reposting rate. The contribution of Tweet is substantial, and the distribution of tweets in the past does not affect the user's reposting rate.

Yang et al. [15] integrated social role identification and information diffusion into an overall framework to model the interaction of users' social roles. Chen etc. [16] studied several semantic features to learn emotional expression tweets.

Mackay, etc. [17] explained in unfamiliar territory that evaluating different prediction models and features can better understand the user's behavior forward. Xu et [18] studied the factors of user posting behavior, including breaking news, posts from users' social friends, and the user's interests, and proposed a potential model to prove these factors' effectiveness further.

Zhang et al. [3] consider users' (Re) Tweet behavior, focusing on how friends in the self-network affect reposting behavior. Unlike existing methods, our method collects image tweets and their titles or comments. Different texts or comments represent pictures with extensive semantic information but are interconnected due to users' social interactions.

**Forwarding Model Representation**:

There has been a lot of research on the modeling of user forwarding representation.
Zhang et al. [1] used a non-parametric statistical model combined with structure, text, and time information to predict forwarding behavior.
Luo et al. [19] developed a learning ranking framework To explore various forwarding characteristics.

Bourgault et al. [20] consider the task from the perspective of time information diffusion. The model learns a diffusion kernel, where the cascading infection time is represented by the node's distance in the projection space.

Jiang et al. [10] proposed A retweet predicting model based on the probability matrix decomposition method is implemented to incorporate observable retweet data, social influence, and message semantics to increase forecast accuracy.

Hoang et al. [9] regard retweet behavior as tweets, tweets with Three-dimensional tensors of the author and his followers, and express them simultaneously through tensor decomposition. On the forwarded data, Bi et al. [21] used two new Bayesian non-parametric models, URM and UCM. The analysis of the tweet text and the user's reposting behavior are integrated into the same probability framework.

Jiang et al. [22]The matrix completion method optimize the user's forwarding representation factorization. Although previous studies have explored learning a wide range of forwarding models that represent users, most of which have not explicitly considered the standard picture of image forwarding and their titles or comments, we have proposed a text guidance Multi-channel network dataset and evaluate its effectiveness using Twitter.

**User retweet Ranking:**

The primary difficulty in retweet prediction is modeling users' tweet-sharing behavior. Users can use the "follower-follower" link to repost tweets and sort all tweets that emerge on social media allowing more users to obtain messages through SMS. This has recently attracted considerable attention in the work of Wang et al. [23].

Liu et al. [24] designed a root function neural network using fuzzy theory, Forwarding behavior modeling. Firdaus et al. [25] proposed a forwarding prediction model based on user authors and forwarding behavior.

Zhang et al. [1] proposed a non-parametric model that combines structural information, text information, and time information to predict forwarding Behavior.

Wang et al. [1] proposed a recommendation model to solve the problems mentioned in tweets by integrating context and social information; this approach employs a deep neural network. A feature-aware factorization model for tweets was proposed by Feng et al. [26]. The model combines the linear discriminant and low-rank factorization models by re-ranking.

Peng et al. [27] explored the partition of social graphs and the building of prediction networks by modeling reposting behavior and using conditional random fields to categorize tweets. The relationship's technique.

## 3. METHODOLOGY

**Image forwarding prediction based on attention ranking network learning**

This part will first explain using a heterogeneous IRM network for image forwarding prediction. Then, based on this, we propose a multi-faceted attention ranking method based on the ranking model of user preference is expressed as $R=\{r_1, r_2, \cdots, r_m\}$ following likes. At the same time, we propose a text-guided multi-modal network that uses the user's context. The attention mechanism guides the image area to represent relevant information jointly.

**Problem definition**

First of all, before explaining the basic model, we need to introduce the basic concepts and professional terms involved in the model. We use $I==\{i_1, i_2, \cdots, i_n\}, \cdots, i_n\}$ Represent a group of image tweets, use $D=\{d_1, d_2, \cdots, d_n\}$. To represent the text content, where $d_i=\{d_{i1}, d_{i2}, \cdots, d_{ik}\}$ Represents the first i.. The text embeddings of different titles and comments for each image tweet. The user set is represented as $U=\{u_1, u_2, \cdots, u_m\}$, The, injuries the first$_i$Users$_{ui}$ The preference characterization embedding. Using the upon symbols, the trouble about picture forwarding reckoning is defined as much follows: Considering the userU or enter photograph twitterI and its associated contextD, Our goal conforms with examine whole user preferencesR. The multi-faceted rating metric characterization about the target user's image tweets are below sorted in conformity with achieving photograph forwarding prediction.

**Heterogeneous IRM network**

To explore the role of image features and contextual information, we use$z=\{With_1, With_2, \cdots, With_n\}$. To represent the joint image Twitter representation. Among them, $With_i I_s$ by the first$_i$ Visual representation of image tweets$_i$ Contextual representation $d_i$. The joint representation composition of the specific fusion method can be seen in a different section. Methods for predicting retweets currently available [1, 3] Only learn user preference models using text tweets that users have previously published and then forecast users' tweet sharing behaviors. Except for earlier research, this paper offers the Heterogeneous IRM network. This multi-dimensional attention ranking measure uses multi-modal image tweets, users' historical forwarding actions, and guiding future to predict image forwarding.

We indicate the proposed heterogeneous IRM network as $G = (V; E)G=(V; E)$, Where joint images represent the node-set V WITH and user preference representation R. Composition, edge set EPast reposting by the userH And its follower relationships S. Composition, using matrix $H \in R^{n \times m}$ Represents the reposting behavior between the image Twitter and the user, where, if the first$_i$ image tweets were ranked$_j$ Users forwarded, then entries$h_{i,j}=1$, Otherwise $h_{i,j}=0$. Then through the matrix$S \in R^{m \times m}$. Consider the following relationship between users, where, if the first$_i$Users follow the $_j$Users, then $s_{ij}=1$. Then use $N_i$ Represents the first$_i$ follow you set of users (if $s_{ij}=1$, Then $u_j \in N_i$), use $N=\{N_1, N_2, \cdots, N_m\}$ represents the user's following collection. After that, the IRM network will be used to derive the heterogeneous triple constraint as the relevant preference for the user to train the multi-faceted attention ranking network.

According to the current Twitter research [28], Twitter shows hidden adverse interest. Set the first$i$ A joint image tweet is represented as With$_{ii}$, NS$_j$ Users are u$_j$, will u$_j$ Non-reposted image twitter of followers With$_k$ Sampling. Via ordered tuples (j, i,k, Nj) Modeling the relative preferences of users, representing the first$j$Users prefer$i$ Image tweets, not images on Twitter. Set T={(j,i,k,Nj)}. A set of nImage tweets and mUsers represents a set of sorted tuples retrieved from the IRM network. The ordered heterogeneous tuples are used as constraints for learning user preference representation or the multi-faceted sorting metric function used to predict image forwarding. For any(j,i,k,Nj)∈T,

The following inequality holds:
$$F_{uj}(With_i) > F_{uj}(With_k) \Leftrightarrow f_{uj}(With_i)hN_j(With_i) > f_{uj}(With_k)hN_j(With_k)n,$$
$$F_{uj}(.) = f_{uj}(.)hN_j(.) F_{uj}(.) = f_{uj}(.)hN_j(.)$$
is used for image forwarding prediction $j$. A multi-faceted sorting model for each user.
$f_{uj}(\cdot)f_{uj}(\cdot)$Function is the first$jj$ Personalized ranking model for each user, $hN_j(.)hN_j(.)$

The model is to follow the preference to the first$j$. The social impact of each user. Under the assumption that the user's preferences may vary according to different parts of the same image tweet, the user's selection. The dimensional space is set as the joint image Twitter representation WITH. The relative ranking module can be used to calculate the user's relative preference result for each image tweet,
where(With$_i$)= rT$_j$With$_i$f$_{uj}$ a personalized sort function, r$_j$I$_s$ the first$_j$.

The relative preferences of each user, With$_i$I$_s$ the first$_i$ A joint representation of image tweets. According to the homogeneity hypothesis [12], individuals in social networks tend to interact and connect with similar others according to their behaviors and opinions. Therefore, we also model the follow-up of users to image tweets Preference for modeling, use(.) to represent.Therefore, the image forwarding prediction problem can be reformulated as follows: Give the joint image Twitter representation WITH, User relative preference T Ordered tuple sets and heterogeneous IRM networks G, Learn all user preferences RI$_s$ represented by a multi-faceted user preference function F u(.) to be sent to the user u. The images are sorted on Twitter.

**Algorithm-1:** Network learning based on text-guided attention ranking
- Step-1:  Initialize I=={i$_1$,i$_2$,⋯,i$_n$},⋯,i$_n$}, T={(j,i,k,Nj)}
- Step-2:  Initialize D= {d$_1$,d$_2$,⋯,d$_n$}, U ={u$_1$,u$_2$,⋯,u$_m$},
- Step-3:  The ranking model of user preference is expressed as R={r$_1$,r$_2$,⋯,r$_m$}, in$r_i$I$_s$ the first$_i$Users$_{ui}$.
- Step-4:  ranking model of user preference is expressed as R={r$_1$,r$_2$,⋯,r$_m$}
- Step-5:  G = (V;E)G=(V;E), Where joint images represent the node-set V WITH and user preference representation R.
- Step-6:  For any(j,i,k,Nj)∈T,V is represented by joint images WITH
    a) Where entriesh$_{i,j}$=1, Otherwiseh$_{i,j}$=0;
    b) Where f$_{uj}$(With$_i$)= rT$_j$With$_i$f$_{uj}$ ;
- Step-7:  matrixS∈R$^{m×m}$ , first$_i$Users follow the $_j$jUsers, then s$_{ij}$=1;
    Then u$_j$∈N$_i$), use N={N$_1$, N$_2$,⋯,N$_m$}
- Step-8:  Set T={(j,i,k,Nj)} For any(j,i,k,Nj)∈T,
- Step-9:  F$_{uj}$(With$_i$)>F$_{uj}$(With$_k$)⇔f$_{uj}$(With$_i$)hN$_j$(With$_i$)>f$_{uj}$(With$_k$)hN$_j$(With$_k$)n,
    F$_{uj}$(.)=f$_{uj}$(.)hN$_j$(.)F$_{uj}$(.)=f$_{uj}$(.)hN$_j$(.)
    Repeat Step-9
    end for
    End

## Network learning based on text-guided attention ranking

In this part, We present a multi-layer multi-faceted interest sorting network that is text-guided. We chose a suitable multi-modal neural community in the IRM network, which consists of two sub-networks: a deep convolutional neural community for a visible illustration of image data and deep recurrent neural networks for text context records semantic representation [29] to symbolize the image tweets. [1] The two sub-networks interact to generate a joint that is defined in the multi-mode integration layer, as in Fig. 2 shown to a group of pictures I={$i_1,i_2,\cdots,i_n$},

First, learn the convolutional features of image Twitter through the last layer of the pre-trained CNN (Convolution neural network) X={$x_1,x_2,\cdots,x_n$}, in =$x_i I_s$ a three-dimensional feature that contains the visual information of the image. Through and F={$f_1,f_2,\cdots,f_n$}. Finally, the same convolutional neural network to learn one wholly visual connection layer embedded image is pre-trained using the initial network. [30] for a visual representation, a visual representation of it in many tasks [31 - 33] widely in Application. Simultaneous training of LSTM (Long short-term memory) network [20] to obtain the relevant context of image tweets, for a set of text contexts D={$d_1,d_2,\cdots,d_n$}.

The latent state of the last hidden layer of the LSTM is used as the semantic embedding of the text context

AND ={$and_1,and_2,\cdots,and_n$} will  {$and_{i1},and_{i2},\cdots,and_{ik}$} represents the semantic embedding of different captions and annotations of image tweets.

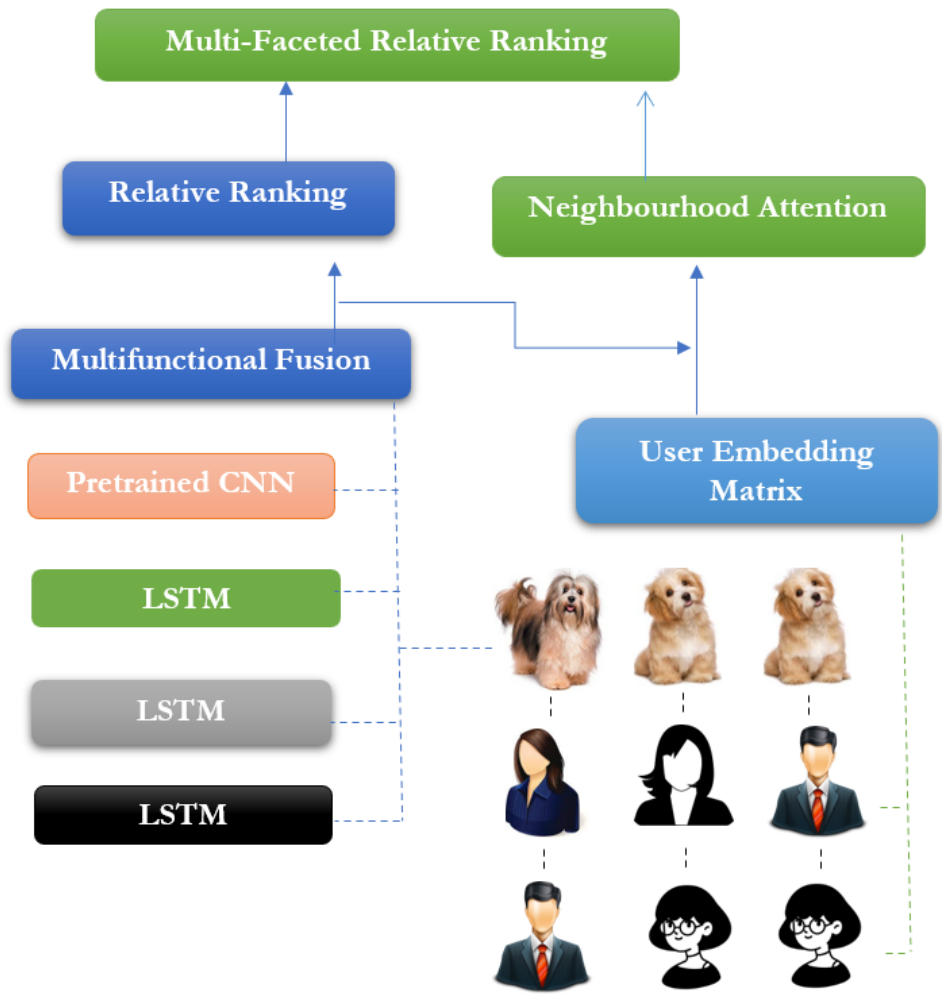

Figure. 2: Network Learning approaching mechanism for multi-faceted attention images or messages sorting used for image feedforwarding prediction with identification and Multi-Faceted relative Ranking System

**Algorithm-2:** Image forwarding prediction based on attention ranking network learning

Step 1: With$_i$ = g ( $W^{(i)} f_i + W^{(d)}$ and $_I$ ),

Where matrix $W^{(i)}$ with $W^{(d)}$ Is the weight matrix($x_i,l_j$)=the($x_i,l_j$)={$the_{i1},the_{i2},\cdots,the_{i9}$} tweak to 9×S9×S

Step2: consider the equation $s_{jk}$=p·tanh($W^{(t)}and^{ij}+W^{(u)} the_{i\ k+b}$)

a) For $a_k$=exp($s_{jk}$)$\sum n \in \eta(x_i,l_j)$,

b) For $g_{ij} = = \sum_k a_k the_{ik}$.
c) Repeat The next position mapping vector is $l_j+1 = g(W^{(j)}f_i + W^{(c)}c_{ij})$
d) ++Add elements of different forms. $W^{(j)}$ with $W^{(c)}$

Step 3: Final calculations: Based on ordered tuple constraints
a) $T = \{(j,i,k,N_j)\}$ calculate $_hN_j(\cdot)$
b) sp $_q = p \cdot tanh(W(s)rp + W(n)rq + b)$
c) $L(j,i,k,N_j) = \max(0, c + F_{uj} - (With_k) - F_{uj} + (With_i))$
d) Where the ranking function
e) $F_{uj}(With_i) = f_{uj}(With_i)hN_j(With_i)$ hyperparameters $c(0<c<1)$

End for
End for
Until forever

## Description

To signify special joint Twitter studying mode image, organizing multi-modal layer, the textual content element of the recurrent neural network representation and visible illustration of the convolutional neural community connecting portion, as in FIG. 2 shown so that the image can be pushed Laid The visual illustration and the semantic representation of the text context are mapped to the identical multi-modal function fusion space. They are delivered to reap the activation of the multi-modal fusion layer, such as. $With_i = g(W^{(i)}f_i + W^{(d)}$ and $_I)$, Where matrix $W^{(i)}$ with $W^{(d)}$ Is the weight matrix, $g(\cdot)$ Is the unsaturated activation function ReLU (Rectified linear unit) [34].

However, this simple method does not utilize the contextual relationship between different comments and their matching image on Twitter. To obtain more relevant image Twitter and text comment representations, this paper proposes a text-guided multi-modal fusion layer, such as FIG 2 shown in detail as Three as shown. Since each image, Twitter has many titles, and comments from their publishers and subscribers may assume a different extension and review information associated with the expression of the picture. Consequently, this Instead of employing the visual features from the pre-trained CNN's last fully connected layer, the picture's convolutional features are used to draw the user's attention to the image.

In this multi-mode fusion network, the attention mechanism is used d to perform Certain constraints simultaneously with text information and, and i Attention mechanisms are implemented to achieve the content of the text information associating the image to it, as Three shown attention modules may be located in an image area adapted to focus a user $L = \{l_0, l_1, \cdots, l_k\}$, $l_i = \{l_{xi}, l_{andi}\}$ Respectively represent the image convolution features $_{xi}$ Shaft and Axis coordinates. Given convolution features $_{xi}$ And position mapping vector $_{lj}$. The convolution in Figure 3 is sampled from $_{xi}$ central $_{lj}$ At $3\times3\times S$ Image features, where SIs the size of the convolution feature. The 3D image feature to be sampled the $(x_i, l_j) = the(x_i, l_j) = \{the_{i1}, the_{i2}, \cdots, the_{i9}\}$ tweak to $9\times S9\times S$. After selecting the convolutional features of the image through the position mapping vector, the attention mechanism is used to fuse the text embedding with the extracted convolutional features, given the first $_i$ Images and i, j Reviews and multi-dimensional features the $(x_i, l_j)$. The semantic features of the can are obtained $_r$eviews and $_k$. The text attention score of each convolution feature is $s_{jk} = p \cdot tanh(W^{(t)}$ and $^{ij} + W^{(u)}$ they $_{k+b})$ in $W^{(t)}$ with $W^{(d)}$ Is the matrix of parameters.

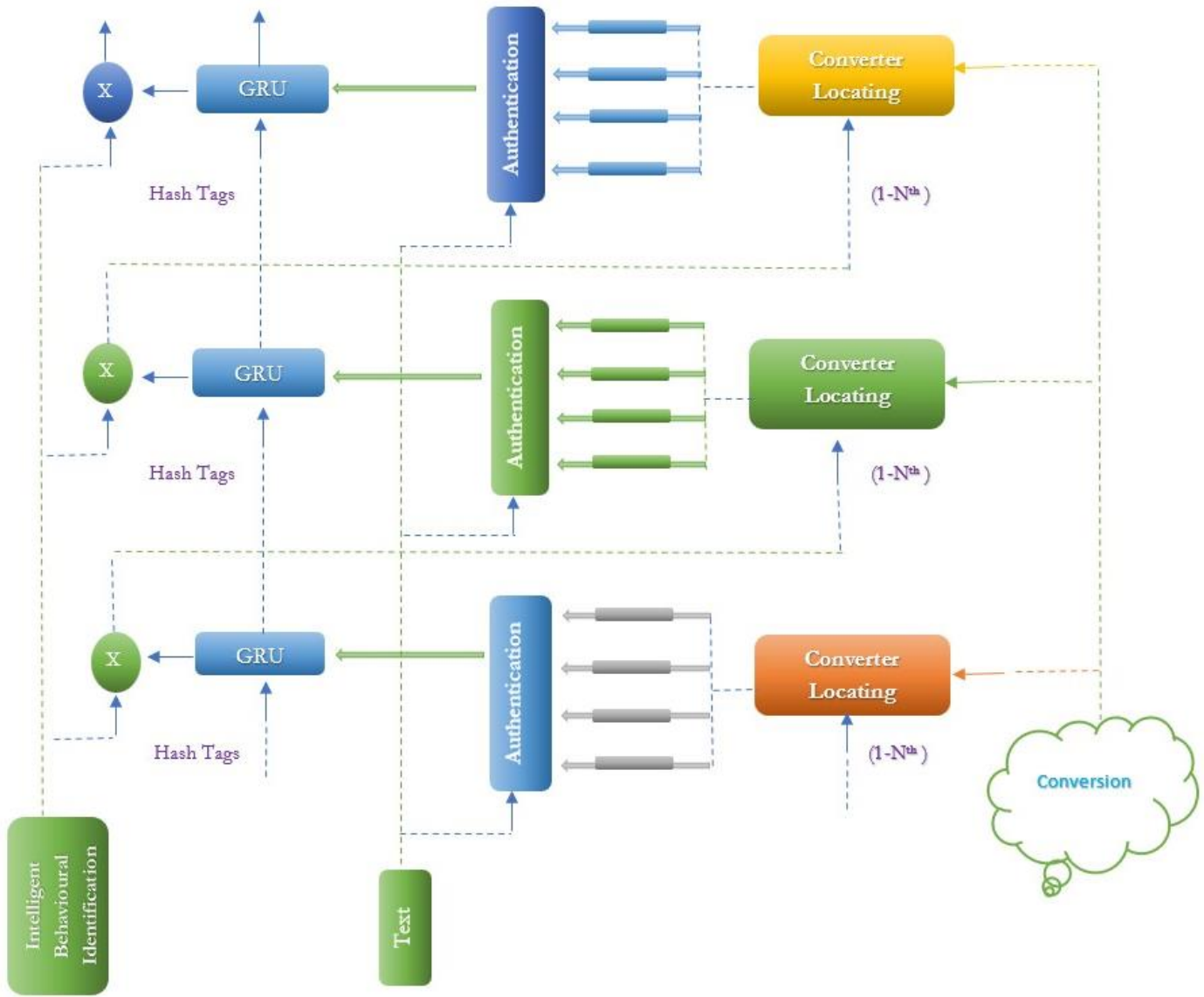

Figure 3  Text and Image Multimode Tweet Hashtag Replicating with a Textually Advanced Ranking System using fusion Network

The partial vector b is used to calculate the text attention score, while the parameter vector pIs are used to calculate the text attention score.

For the $(x_i,l_j)$ Each follower in $th_{ek}$, Its score is activated as $a_k=\exp(s_{jk})\sum n\in \eta(x_i,l_j)$, And the text is$_j$. The influence of image convolution features is $g_{ij}= =\sum k a_k the_{ik}$.

Next, use another recurrent neural network to infer the location of the following image region.$g_{ij}g_{ij}$As the first$_j$ Time step input, use$h_{ij}$with$_{cij}$ Represents the hidden state and output of RNN (Recurrent neural networks), The visual elements come from the pre-trained CNN's last ultimately linked layer.

The visual embedding of a given image$_{fi}$ And RN's$_{jj}$ Step output$_{cij}$,
The next position mapping vector is $l_j+1=g(W^{(j)}f_i+W^{(c)}c_{ij})$ ++Add elements of different forms. $W^{(j)}$ with $W^{(c)}$ Is the weight matrix. $g(\cdot)g(\cdot)$Is the hyperbolic tangent function scaled by the unit. Definition TG ($l_j,x_i,$and$_{ij},f_i$).

For the text guidance process, by superimposing the model with the recurrent neural network, the next position mapping vector and the hidden state of the RNN can be obtained.

On this basis, a multi-faceted attention ranking function learning method is proposed for image forwarding prediction. Inspired by the attention mechanism [1], the neighborhood attention module is designed in Figure 2, which has a social impact function $_hN_j(\cdot)$ Perform calculations.

Based on ordered tuple constraints $T=\{(j,i,k,N_j)\}$ calculate $_hN_j(\cdot)$. Given user preference representation $R=\{r_1,r_2,\cdots,r_n\}$, The social influence attention scores of users and their followers are expressed as

$$sp_q = p \cdot tanh(W^{(s)}r_p + W^{(n)}r_q + b)$$

$nW^{(s)}$) with $W^{(n)}$ Is the parameter matrix. b Is the partial vector, p It is the parameter vector that calculates the social influence attention score.

Finally, we can define the attention multi-faceted ranking loss function as follows:

$$L(j,i,k,N_j) = \max(0, c + F_{uj}-(With_k) - F_{uj}+(With_i))$$

Where the ranking function $F_{uj}(With_i) = f_{uj}(With_i)hN_j(With_i)$, Superscript $F+_{uj}(\cdot)$ Indicates a favorable preference, $F_{uj}-(\cdot)$ Represents negative preferences, hyperparameters $c(0<c<1)$ Control the boundaries of the loss function.

## Data set Preparation

**Information about the data set:** We gathered data from Twitter, a popular social networking service that allows users to circulate their broadcasting material. [36]. Users habitually repost pictures on social networking media platforms to show that they push images with Special positive preferences.

We capture the user's personal information based on open social network platforms, with the image or text tweets they have reposted in the past and their following subscriptions through maintaining and working under chain relationships.

It is known that the spread of information tends to decay with distance and the user's interest index. Image reposting is a cascade of information A forms presenting a heavy-tailed distribution in the user relationship network.

a) The data set is crawled in the "Twitter-User-Twitter" cycle to avoid this phenomenon as much as possible. In other words, we first choose a repost For Weibo more than five times find its publisher. Then, we collect the image tweets of the publisher in the last month and find users who repost these tweets.
b) Users with less than eight followers are filtered, and then We use the rest of the users to find their pictures on Twitter again.
c) We loop the "Twitter Users" 5 times, extract all followers and followers, and build an image forwarding modeling network. We collected 15,500 users, 74,927 image tweets, and 274,851 follow relationships.
d) Image feature extraction: We conduct the following preprocessing on the gathered image tweets.
e) For picture feature embedding, we extract global features from the last fully connected layer of the pre-trained Inception-V4 network, a total of 1536 dimensional vectors.
f) We additionally extract picture characteristics from the last convolutional layer of the same pre-trained network to obtain 8x8x1536 feature vectors for each image to suit the needs of text-guided multi-modal networks.
g) Text feature extraction: We advance entire filter emojis yet interjections to achieve whole titles then comments.
h) Then, we utilize the pre-trained Glove model [37] to expel semantic representation for every sentence word. The dimension of word vectors
i) The number is 300. Expressly, four sentences are set for each image tweet, and the length of each sentence is 12.
j) We copy the last comment as padding for image tweets with less than four headlines or comments. For our data set, the vocabulary size is set to 12500. Therefore, we use the word tags and <eos> to mark the end of the title or comment.

## Evaluation Criteria

In most online media services, forwarding prediction tasks are designed to provide users with$_K$ Image push. To evaluate our method in the top ranking$_s$ for the effectiveness of Twitter image, we use two ranking-based evaluation criteria, exactness [2] and ROC (Receiver operating characteristic curve). The area under the curve and the coordinate axis (Area under curve, AUC) [38 is - 40]. to evaluate the image transfer performance prediction given user$U_tU_t$ And image twitter$_{it}$ Test set, use $R_{ui}$ Indicates a user in the test set$_{ui}$ Before$_K$ Predicted rankings of image Twitter, where the ranking list $|R_{ui}|$ The size is K.

## Effect Comparison

Evaluate the performance of the methods in this article, AMNL (only linear fusion method) and AMNL+ (multi-modal network guided by text), and five other latest solutions to the image forwarding prediction problem.

Table 1  Experimental results on accuracy of changed methods

| Method | Accuracy@1 | | |
|---|---|---|---|
| | 60 % | 70 % | 80 % |
| RRFM | 0.61253 | 0.64743 | 0.65383 |
| VBPR | 0.63399 | 0.65211 | 0.67931 |
| D-RNN | 0.70001 | 0.71910 | 0.73851 |
| CERBERUS | 0.71930 | 0.72951 | 0.75161 |
| ADABPR | 0.63942 | 0.64881 | 0.66921 |
| CITING | 0.74630 | 0.76080 | 0.77730 |
| AMNL | 0.86910 | 0.89751 | 0.90080 |
| **AMNL+** | **0.93418** | **0.94446** | **0.95857** |

Table 2  Experimental results on precision@3 of different approaches

| Method | Accuracy 3 | | |
|---|---|---|---|
| | 60 % | 70 % | 80 % |
| RRFM | 0.5973 | 0.6284 | 0.6400 |
| VBPR | 0.6082 | 0.6304 | 0.6432 |
| D-RNN | 0.6468 | 0.6702 | 0.6879 |
| CERBERUS | 0.6593 | 0.6684 | 0.6813 |
| ADABPR | 0.5980 | 0.6198 | 0.6301 |
| CITING | 0.7304 | 0.7467 | 0.7677 |
| AMNL | 0.7519 | 0.7791 | 0.7959 |
| **AMNL+** | **0.8680** | **0.8796** | **0.8823** |

Table 3  Experimental results on AUC of different approaches

| Method | AUC | | |
|---|---|---|---|
| | 60 % | 70 % | 80 % |
| RRFM | 0.5032 | 0.5195 | 0.5282 |
| VBPR | 0.5491 | 0.5799 | 0.5814 |
| D-RNN | 0.6834 | 0.6973 | 0.6999 |
| CERBERUS | 0.7145 | 0.7342 | 0.7440 |
| ADABPR | 0.5393 | 0.5601 | 0.5782 |
| CITING | 0.5802 | 0.5982 | 0.6425 |
| AMNL | 0.7703 | 0.7998 | 0.8486 |
| **AMNL+** | **0.8792** | **0.8986** | **0.9126** |

Table 1, Table 2, and Table 3 respectively show the outcomes of all techniques' evaluations based on the ranking criteria Precision@1, Precision@3, and AUC[1]. This article employs three data sets as the training set for assessment: 60%, 7%, and 80% of the data. The grade evaluation standard compares all of the techniques' outcome values. Then, examine the model's performance in various settings, where the user preference dimension is set to 400, and 80 percent of the data is used for training. All other parameters and hyperparameters are also selected to ensure.

The best performance on the validation set. We evaluate the average of all three criteria on six methods. These experimental results reveal some interesting points:
   a) CITING, D-RNN, and VBPR outperform the low-rank parametric ranking metrics ADABPR and RRFM when using content features as auxiliary information for learning ranking metrics, demonstrating that the deep neural network contains both images push and related context information. It is critical to the image forwarding prediction problem.
   b) Compared with other sorting methods that contain side information, $AMNL_I$ has better performance than VBPR, and $AMNL_d$ has better performance than CITING. This shows that multi-faceted ranking indicators are essential.
   c) Compared to the AMNL approach, the AMNL+ method performs better. This demonstrates that the text-guided multi-modal fusion algorithm may more effectively merge picture tweets with varied titles or comments containing relevant semantic information, improving image forwarding prediction performance.
   d) In all circumstances, the AMNL+ technique provides the best results. This demonstrates that the multi-modal picture Twitter representation and its related context and the multi-dimensional ranking measure attention multi-dimensional ranking network learning framework may be improved. Improve image forwarding prediction performance.

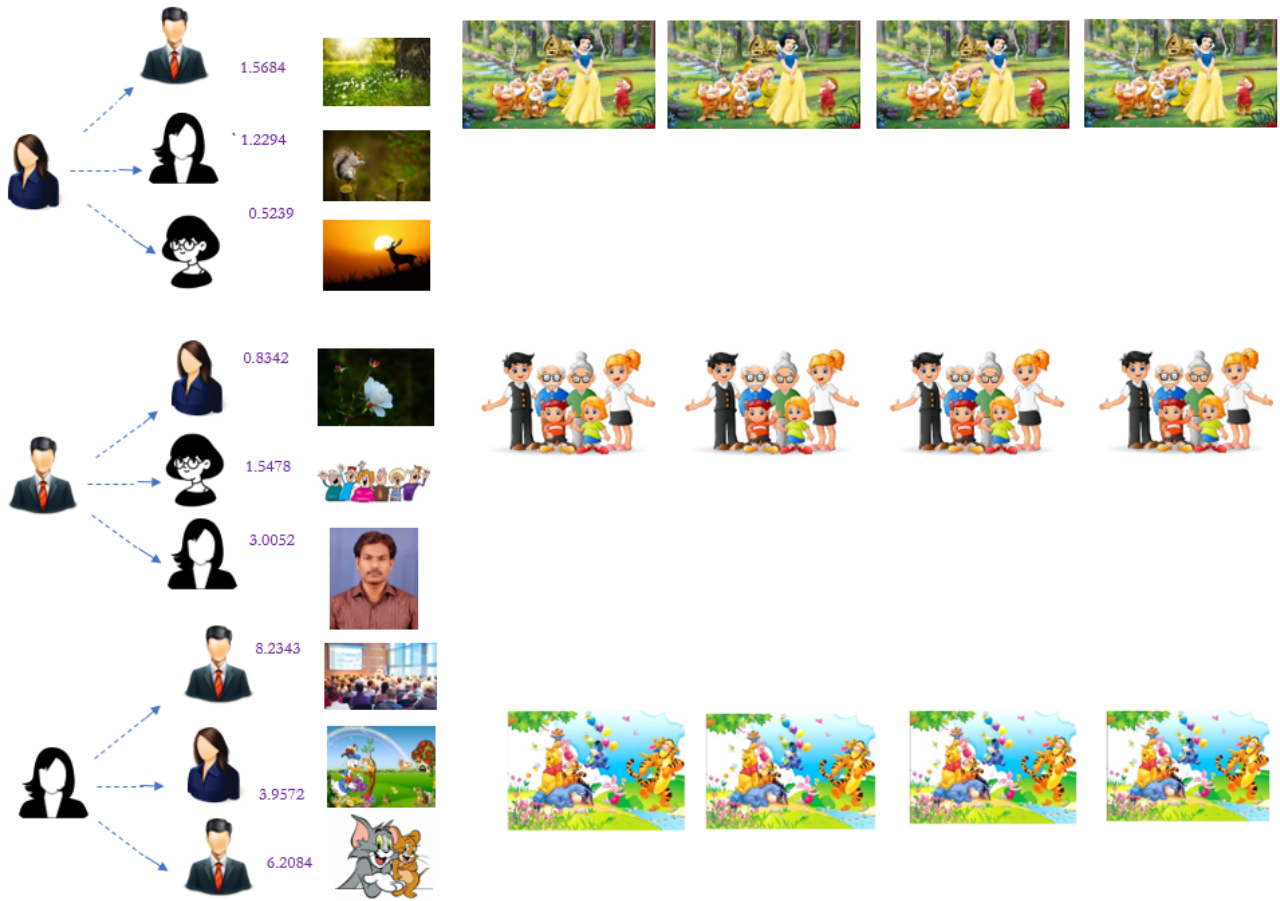

Figure 4   Experimental results of AMNL+ and Advanced ranking System using fusion Network in the image forwarding prediction mechanism

Figures 4 show examples of AMNL+'s experimental results for some users' image forwarding predictions. We mainly select sentence combinations with different degrees of relevance to image tweets and grammatical and semantic complexity in text selection. For example, there are different objects, etc., to test the actual utility of the attention mechanism. Moreover, the recommended tweets often have more exciting and relatively complex semantic content in natural phenomena.

Due to space constraints, consider three sentences in the example shown in Figure 4. Figure 4 is divided into two sections:
(a). the image posted by the user and the user's followers, the ranking score predicted by the model, and
(b). the feedback effect of the predicted Twitter image and its annotations on the attention module. Figure 4 shows the predicted likes of various Twitter accounts, and the lower the ranking score, the less likely it is to be recommended.

We can find that the attention is The unreported image tweets posted by the author have obtained poor ranking scores. This demonstrates that while these tweets are more likely to be seen by users, they are not widespread and consistent with the facts. Figure 4 shows several examples of high annotation scores in image tweets, and the keywords in other annotation comments have been established with good attention in the image. This indicates that the idea predicted by our method is better than in Figure 4. The user is desirable. It is worth mentioning that some specific words are matched with objects of the same color in the image, which reflects that annotations and subtitles have a particular guiding effect on prediction. The training update rules of the attention multi-faceted ranking network learning method proposed in this paper are essentially iterative. Next, this paper will continue to study the convergence of the AMNL method.

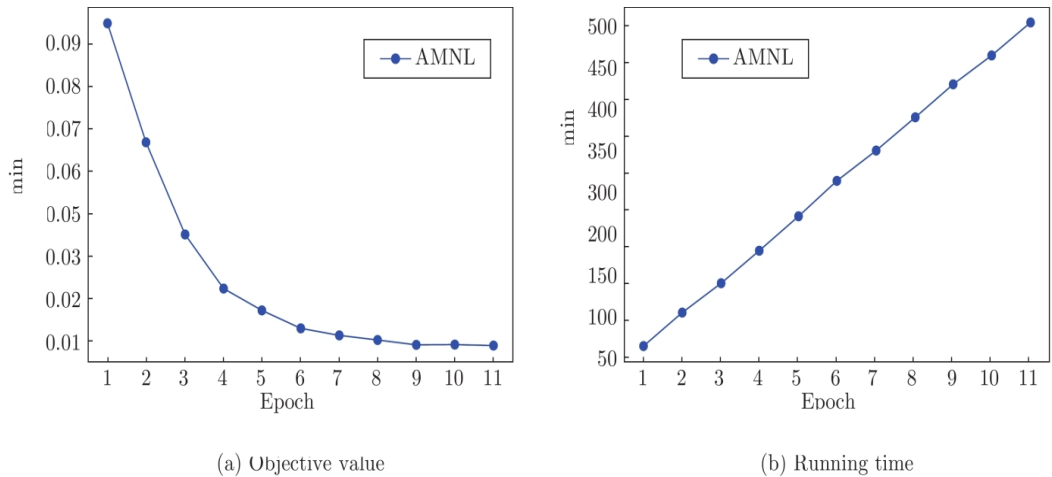

(a) Objective value       (b) Running time

Figure 5    Results changes with the objective value and running time of Epoch based on Images and Texts

Figures 5(a) and 5(b) are the convergence of the AMNL method. And run-time curve graph. FIG $x$-axis represents the number of iterations. Figure 5(a) in the $y$-axis target value, Figure 5(a) in the $y$-axis is the operating time of the proposed approach—each Epoch 231,539 comprising Iterative update. We made the person preference illustration a size of 500 and used 80% of the information for training. The results exhibit that the method completes convergence after ten rounds, and the calculation time is about 500 minutes. This lookup validates AMNL's Effectiveness.

## Ablation Experiment

The contribution of our model module was evaluated, with a particular focus on the text-guided multi-modal fusion community and social effect function. We tested the model's visual representation of picture tweets, semantic illustration of associated contexts, and joint photo tweet illustration all at the same time. We propose an ablation study and list the model to understand the contribution of components and the influence of different media on our model: We only use the visual representation of image tweets in the AMNL I method. The AMNL d method implies that only the relevant context is semantically represented. The AMNL $_{+i}$ model is in a text-guided multi-modal fusion network. The average pooling of image Twitter convolution features is directly input into the recurrent neural network instead of using the attention mechanism in the text representation. AMNL $_{hfunc}$ and The AMNL $_{+ hfunc}$ model means that we directly calculate the ranking functions of the two models without using the social influence function. As shown in Table 4, we also found some interesting results:
a)  Compared with the AMNL i and AMNL d methods, the AMNL approach achieves higher performance.
b)  This suggests that in contrast to only the use of visual points or textual content features, the interest in multi-faceted rating networks gaining knowledge of framework uses the joint image of multi-modal photos Twitter characterization and its associated context can achieve better performance.

c) AMNL $_{+ hfunc}$, AMNL $_+$ scores higher in the three criteria. This suggests that the social impact function can aid in the improvement of our method's performance. The experimental results of AMNL $_{+ hfunc}$ and AMNL further prove our the above results are consistent between different components.

Table 4: Results of experiments with various modalities and components using 80% of the data for training

| Method | Precision@1 | Precision@3 | AUC |
|---|---|---|---|
| AMNL$_{+i}$ | 0.8467 | 0.7673 | 0.8204 |
| AMNL $_d$ | 0.7982 | 0.7719 | 0.7962 |
| AMNL $_{hfunc}$ | 0.8698 | 0.7900 | 0.8095 |
| AMNL | 0.9009 | 0.7959 | 0.8486 |
| AMNL$_{+i}$ | 0.9227 | 0.8276 | 0.8724 |
| AMNL$_{+hfunc}$ | 0.9199 | 0.8195 | 0.8689 |
| **AMNL+** | **0.9585** | **0.8823** | **0.9126** |

## 4. CONCLUSION

This paper proposes an image forwarding prediction model based on a heterogeneous IRM network. Specifically, our IRM network uses the image tweets forwarded by the user in the past, the associated text context, and the user's subsequent relationship to sample an appropriate representation of the user's forwarding behavior. On this basis, we propose a text-guided multi-modal neural network attention multi-faceted ranking method to learn joint image Twitter representation and user preference representation, thereby embedding multi-faceted ranking metrics into the picture. Make predictions. The performance of our technique is evaluated using the Twitter dataset. A vast number of tests have revealed that our strategy outperforms several of the most current methods.

**Competing Interests**

All Authors declare that they have no competing interests.